# The Circle of Investment

*Connecting the Dots of the Portfolio Management Cycle … under the Purview of the Uncertainty Principle of the Social Sciences*

Edited Version: Kashyap, R. "The Circle of Investment." International Journal of Economics and Finance, Vol. 6, No. 5 (2014), pp. 244-263.

Now expanded into a book titled, The Circle of Investment, available on www.Amazon.com.

**Ravi Kashyap**
Gain Knowledge Group
City University of Hong Kong
May 9, 2013

## Table of Contents

# I. Abstract

*The Circle of Investment: Connecting the Dots of the Portfolio Management Cycle - Hypothesis Formulation; Portfolio Construction; Trade Execution; Risk Management; Performance Measurement and Portfolio Rebalancing - under the Purview of the Uncertainty Principle of the Social Sciences.*

We will look at the entire cycle of the investment process relating to all aspects of, formulating an investment hypothesis; constructing a portfolio based on that; executing the trades to implement it; on-going risk management; periodically measuring the performance of the portfolio; and rebalancing the portfolio either due to an increase in the risk parameters or due to a deviation from the intended asset allocation.

We touch upon the fundamentals of multi-factor models and how they are used across the different stages of the investment cycle. We also provide *several illustrative analogies that are meant to intuitively explain the pleasures and the pitfalls that can arise while managing a portfolio.*

If we consider the entire investment management procedure as being akin to connecting the dots of a circle, then *the Circle of Investment can be represented as a dotted circle with many dots falling approximately on the circumference and with no clue about the exact location of the centre or the length of the radius.*

*We represent the investment process as a dotted circle since there is a lot of ambiguity in the various steps involved. The circle also indicates the repetitive nature of many steps that are continuously carried out while investing.*

While there are numerous methods that can be applied to each step of the process, we mention the ones that are most used in practice and highlight the elements that a practitioner needs to watch out for. In the beginning, we consider the idea of market efficiency and equilibrium and the lack of both, though we find that there is a tendency to move towards efficiency and the establishment of states of pseudo-equilibrium. This leads to the realization that any hypothesis comes with limitations and that investments are constantly under the shadow of this uncertainty.

This work introduces two new points pertaining to this dotted circle and improves the ability; to understand how far-off this dotted circle is, from a more well-defined circle and; to create a well-formed circle. One point lies close to the centre of the circle and helps clarify both the size and shape of the circle. The other point lies on the periphery of the circle and helps with forming a more round shape.

*The two innovations we introduce regarding the investment life-cycle are:*

1. *The first, relating to the limitations that apply to any finding in the social sciences, would be the additional point we introduce that lies near the centre of the circle. We title this as, "The Uncertainty Principle of the Social Sciences".*
2. *The second, relating to establishing confidence levels in a systematic manner for each view we associate with a security or group of securities as required by the Black Litterman framework, would be the new point we present near the circumference of the circle.*

We restrict ourselves primarily to the equity asset class, while clarifying earlier on that the main differences between asset classes are simply due to the contractual terms and the number of parties involved in the transfer of wealth. In addition to equities, we look at the execution costs that apply to foreign exchange, fixed income and commodities. This is important since some equity portfolios could be across different markets and hence have currency exposure; or the portfolio could hold high grade fixed income instruments, in lieu of holding cash; or there might be the occasional active bet on commodities to increase the return or as a diversification measure.

## II.    Uncertainty Principle of the Social Sciences

At the outset, let us look at some fundamentals that govern all financial instruments and then delve into the nuances for the Equity asset class. It is also worthwhile to mention here that for most assertions made below, numerous counter examples and alternate hypothesis can be produced. These are strictly attempts at tracing the essentials rather than getting bogged down with a specific instance. However, investing requires forming a conceptual framework based on the more common observations, yet being highly attuned to any specifics that can stray from the usual. Also, for the sake of brevity, a number of finer points have been omitted and certain simplifying assumptions have been made.

The various financial instruments that exist today can be broadly viewed upon as vehicles for providing credit and a storage for wealth, for both individuals and institutions alike. The different instruments, both in terms of their nomenclature and their properties, then merely become manifestations of which and how many parties are involved in a transaction and the contractual circumstances or the legal clauses that govern the transaction.

Despite the several advances in the social sciences and in particular economic and financial theory, *we have yet to discover an objective measuring stick of value, a so called, True Value Theory*. While some would compare the search for such a theory, to the medieval alchemists' obsession with turning everything into gold, for our present purposes, the lack of such an objective measure means that the difference in value as assessed by different participants can effect a transfer of wealth. This forms the core principle that governs all commerce that is not for immediate consumption in general, and also applies specifically to all investment related traffic which forms a great portion of the financial services industry.

Although, some of this is true for consumption assets; because *the consumption ability of individuals and organizations is limited and their investment ability is not*, the lack of an objective measure of value affects investment assets in a greater way and hence investment assets and related transactions form a much greater proportion of the financial services industry. Consumption assets do not get bought and sold, to an inordinate extent, due to fluctuating prices, whereas investment assets will. The price effect on consumptions assets affects the quantity bought and consumed, whilst with investment assets, the cyclical linkage between vacillating prices and increasing number of transactions becomes more apparent.

***Another distinguishing feature of investment assets is the existence or the open visibility of bid and ask prices***. Any market maker for investment assets quotes two prices, one at which he is willing to buy and one at which he is willing to sell. Consumption assets either lack such an outright two sided quote; or it is hard to painlessly infer viewable buy and sell prices, since it involves some conversion from a more basic form of the product into the final commodity being presented to consumers. Examples for consumption assets are a mug of hot coffee, that requires a certain amount of processing from other rudimentary materials before it can be consumed; or a pack of raw almonds which is almost fit for eating. Coffee shops that sell coffee do not quote a price at which they buy ready drinkable coffee; the price at which a merchant will buy almonds is not readily transparent. Gold is an example of both, a consumption and an investment asset. A jewellery store will sell gold and objects made of gold; but it will also buy gold reflecting its combined consumption and investment trait. This leaves us with financial securities like stocks and bonds that are purely investment assets.

A number of disparate ingredients contribute to this price effect; like how soon the product expires and the frequent use of technology to facilitate a marketplace. EBay is an example of a business where certain consumption goods are being bought and sold. This can happen even if goods are only being sold, through the increased application of technology in the sales process. While not implying that the use of technology is bad, technology, or almost anything else, can be put to use that is bad. Thankfully, we are not at a stage where Starbucks will buy and sell coffee, since it can possibly lead to certain times of the day when it can be cheaper to have a cup of coffee and as people become wary of this, there can be changes to their buying habits, with the outcome that the time for getting a bargain can be constantly changing; making the joys of sipping coffee, a serious decision making affair. Even though this is an extreme example, we will overlook some of these diverse influences for now, since our attempt is to exemplify the principal differences between the varieties of financial transactions and the underlying types of assets that drive these deals.

This lack of an objective measure of value, (henceforth, value will be synonymously referred to as the price of an instrument), makes prices react at varying degrees and at varying speeds to the pull of different macro and micro factors. The greater the level of prevalence of a particular instrument (or even a particular facet of an instrument) the more easily it is affected by macro factors. This also means that policies are enforced by centralized institutions, (either directly by the government or by institutions acting under the directive of a single government or a coalition of governments), to regulate the impact of various factors on such popular instruments. Examples for this would be interest rate dependent instruments, which are extremely sensitive to rates set by central banks since even governments issue such instruments; dividends paid by equity instruments which are clearly more sensitive to the explicit taxation laws that govern dividends than to the level of interest rates; and commodities like oil, which are absolutely critical for the smooth functioning of any modern society and hence governments intervene directly to build up supplies and attempt to control the price.

Lastly, it is important that we lay down some basics regarding the efficiency of markets and the equilibrium of prices. Surely, a lot of social science principles and methodologies are inspired from similar counterparts in the natural sciences. A central aspect of our lives is uncertainty and our struggle to overcome it. Over the years, it seems that we have found ways to understand the uncertainty in the natural world by postulating numerous physical laws. These physical laws are deductive and are based on three statements - a specific set of initial conditions, a specific set of final conditions and universally valid generalizations. Combining a set of generalizations with known initial conditions yields predictions; combining them with known final conditions yields explanations; and matching known initial with known final conditions serves as a test of the generalizations involved. The majority of the predictions in the physical world hold under a fairly robust set of circumstances and cannot be influenced by the person making the observation and they remain unaffected if more people become aware of such a possibility.

In the social sciences, the situation is exactly the contrary. A set of initial conditions yielding a prediction based on some generalization, ceases to hold, as soon as many participants become aware of this situation and act to take advantage of this situation. This means that predictions in the social sciences are valid only for a limited amount of time and we cannot be sure about the length of this time, since we need to factor in constantly the actions of everyone that can potentially influence a prediction, making it an extremely hard task.

***All attempts at prediction, including both the physical and the social sciences, are like driving cars with the front windows blackened out*** and using the rear view mirrors, that give an indication of what type of path has been encountered and using this information to forecast, what might be the most likely type of terrain that lies ahead for us to traverse. The path that has been travelled then becomes historical data that has been collected through observation and we make estimates on the future topography based on this. Best results generally occur, when we combine the data we get in the rear view mirror with the data we get from the side windows, which is the gauge of the landscape we are in now, to get a better comprehension of what lies ahead for us. The quality of the data we gather and what the past and the present hold then give an indication to what the future might be. So if the path we

have treaded is rocky, then the chances of it being a bumpy ride ahead are higher. If it has been smooth, then it will be mostly smooth. Surely, the better our predictions, the faster we can move; but then again, it is easy to see that the faster we travel, the more risk we are exposed to, in terms of accidents happening, if the constitution of the unseen scenery in front of us shifts drastically and without much warning.

A paramount peculiarity of the social sciences is that passage on this avenue is part journey and part race. The roads are muddy, rocky and more prone to have potholes. This means being early or ahead on the road brings more winnings. We also have no easy way of knowing how many people are traveling on this path, either with us, ahead of us or even after us. As more people travel on the path, it starts falling apart, making it harder to travel on it, a situation which is accentuated considering we don't have any vision out front. On the other hand, let us say, physical science roads, being well paved and well-constructed using concrete, hold steady for much longer time durations, so what has been observed in the past can be used to make durable forecasts that hold for lengthier amounts of time in the future.

This inability to make consistent predictions in the social sciences and the lack of an objective measure of value or a True Price Theory means that is almost impossible for someone to know what a real state of equilibrium is. The efficient market hypothesis in spite of being a very intriguing proposition, can at best claim that markets have a tendency to move towards being efficient, though a state of equilibrium is never fully attained since no one has an idea what that state of equilibrium is and the actions of the participants serves only to displace any state of equilibrium, if it did exist. The analogy for this would be a pendulum with perpetual motion; it swings back and forth around its place of rest with decreasing amplitude and the place of rest keeps changing with time, starting a new cycle of movement with reinforced vigour.

We can then summarize the above with the *Uncertainty Principle of the Social Sciences*,

which can be stated as, "***Any generalization in the social sciences cannot be both popular and continue to yield predictions or in other words, the more popular a particular generalization, the less accurate will be the predictions it yields***". This is because as soon as any generalization and its set of conditions become common knowledge, the entry of many participants shifts the equilibrium or the dynamics, such that the generalisation no longer applies to the known set of conditions.

All our efforts as professionals in the field of investment, will then be to study uncertainty and uncover quasi-generalizations; understand its limitations in terms of what can be the closest states of pseudo-equilibrium; how long can such a situation exist; what factors can tip the balance to another state of temporary equilibrium; how many other participants are aware of this; what is their behaviour and how is that changing; etc., making our professions a very interesting, challenging and satisfying career proposition.

With this in mind, we can turn specifically to how the above discussion applies to the Equity Asset class.

## III.     Important Elements of the Equity Asset Class

- The Equity asset class holds the potential for unlimited upside and brings with it partial ownership of the firm and hence some influence over the decision making process. It can be argued that ***this premise of boundless profits, coupled with limited losses or liability and a certain degree of control, make this asset class an extremely appealing one***, contributing to its immense popularity.
- Most equity instruments are traded on exchanges and the act of listing itself serves as a signal of confidence to potential investors or the public. Trading on exchanges also means the counterparties are anonymous as opposed to fixed income and FX markets where a lot of deals are done on the phone.

- The debt of two similar companies will be more identical to each other than the equity of the same two companies (This is purely in terms of how sensitive the instruments are to various stimuli), making the equity asset class the most granular in terms of the number of different types of instruments and markets around the world.
- The overall size of the equity asset class is smaller than the fixed income and FX markets.
- While, Market Efficiency is non-existent in its strictest sense in almost every asset class, the Equity asset class has the least tendency to be efficient since it lacks any underpinning forces or levers that can serve as constraints for the establishment of equilibrium. Commodities have the limited supply of some physical product acting as a controlling lever; fixed income and FX instruments have interest rates that are artificially set; but the closest thing that equities have to determine prices are its expected dividends which are extremely prone to fluctuations both within the same instrument and certainly across instruments. (It is a much longer and harder discussion as to what the level of interest rates should be. It is generally accepted that supply and demand conditions cause the violent fluctuations seen in commodity prices).
- Randomness or noise exists to a greater extent since there are numerous participants and less forces or levers that can be used to control price levels as opposed to other financial instruments like fixed income, commodities and FX.
- ***The various asset classes can be compared to balloons tethered to the ground, with the equity balloon having the weakest connections to the ground and also the weakest controls to guide it, if it is wind-borne***.
- The lack of a strong controlling factor also makes regime changes much harder to detect. Regime changes are a major shift in the investment landscape, or from our

Ravi Kashyap                                                                                                                    Page 10

earlier analogy, this would be a change in the resting place of the pendulum.

- The behaviour of equity prices are modelled as Brownian motions with a certain amount of drift, which is usually the rate of return of the instrument. Historically, this model seems to work fairly well since equity prices have been known to increase over time with a certain amount of noise or variation around that long term growth rate.

- ***Equity prices are considered to be Markovian***, that is historical prices have no ability to predict future prices. While other asset classes are also Markovian to a great extent, equity prices are more Markovian than the rest since they are subject to a greater amount of randomness and lack strong controlling factors.

- Generally, falling equity prices are a leading indicator of economic contraction and high equity prices are lagging indicators of an asset price bubble build up.

- The equity asset class has seen a large number of bubbles since its inception and it seems to be a periodically recurring phenomenon. This again can be partly attributed to the lack of any major controlling factor over equity prices. It is possible to separate the formation and bursting of bubbles into five different stages.

  - Displacement – Some change in economic circumstances creates new and profitable opportunities for certain companies.
  - Euphoria – In this stage, the growth prospects of the companies or the expected profits are vastly overestimated and lead to rapid price growth.
  - Mania – Many first time investors enter the market seeking to make quick capital gain returns. This is also referred to sometimes as a herding phenomenon, where people do things because others are doing it.
  - Distress – The early entrants or the more savvy investors see that the expected profits are not justified and cash out with their profits.
  - Revulsion – The market begins to fall and causes a stampede of investors to

pull out their money, resulting in a number of investors facing severe losses.

- After the market crash of 1987, the equity markets have started displaying a positive skew towards lower prices. This means that the ***probability of a huge downward move in prices is more than the chance of an upward move of similar magnitude***.

- Historically, ***equities have outperformed most other investments over the long run***. This is attributed to the slightly higher risk associated with Equity investments since in the event of a company going bust, shareholders are the last group that has any claim on the assets of the firm.

- The percentage of house hold savings being directed to equities has increased over the last few decades. Though it has dipped after the most recent financial crisis, the equity share is expected to make a resurgence.

- Last but not the least; its origins can be traced to the Netherlands.

## IV. Characteristics of a Good or a Bad Trade

- The factors that dictate a good trade or a bad trade depend on the Time Horizon and the Investment Objective. The time horizon can be classified into short term, medium term and long term. The investment objective can be conservative or aggressive. While there are no strict boundaries between these categories, such a classification helps us with the analysis and better identification of trades.

- ***Any trade that fulfils the investment objective and time horizon for which it is made is a good trade. Otherwise, it is a bad trade***.

- On the face of it, we can view good trades as the profitable ones and bad trades as ones that lose money. But where possible, if we try and distinguish between proximate causes and ultimate reasons, it becomes apparent that ***good trades can lose money and bad trades can end up making money***.

- As discussed in the introduction, the noise around the expected performance of any security; our ignorance of the true equilibrium; the behaviour of other participants; risk constraints (these will be discussed in the later sections) like liquidity, concentration, unfavourable geo-political events; etc. implies we would have deviations from our intended results. The larger the deviation from the intended results, the worse our trade is.

- What the above implies is that, bad trades show the deficiencies in our planning (estimation process) and how we have not been able to take into account factors that can lead our results astray. It is true that due to the extreme complexity of the financial markets, the unexpected ends up happening and we can never take into account everything. ***We just need to make sure that the unexpected, even if it does happen, is contained in the harm it can cause***. The good thing about bad trades is the extremely valuable lessons they hold for us.

- We then need to consider how a good trade can lose money. When we make a trade, if we know the extent to which we can lose, when this loss can occur and that situation ends up happening, our planning did reveal the possibility and extent of the loss, hence it is a good trade.

- The bottom line is that, good trades or bad trades are the result of our ability to come up with possible scenarios and how likely we think they will happen.

- The following are some other factors that can contribute to good equity trades.

- The trade will not soak too much of the available liquidity, as measured by the average trading volume, unless of course, we wish to take a controlling stake in the firm.

- It is held by a number of investors. There is more uncertainty if there are more investors, but it seems to work to our benefit in most cases. If the number of investors is limited, the possibility of all of them doing the opposite of what we want is higher and more likely.

- The noise or the randomness is less so that our decisions can be more accurate. This can be measured by volatility or the price fluctuations that we see.

- The firm issuing the securities is not too dependent on any particular product, profits from a particular region, is not overburdened with debt, is paying dividends consistently, its price is not too high compared to its earnings and other fundamental research indicators.

- If we are able to see some pattern in the share price changes, that is a good trade. This means that this security is exhibiting non-Markovian behaviour. Such behaviour is usually hard to detect, but it comes down to the lens we are using to view the world or the methods we are using to perform historical analysis.

- If the security is affected by any asset price bubbles and we are able to detect the formation of such bubbles.

- If we are shorting the security and it has a greater tendency for a downward movement, as exhibited by its skew.

## V.  Main Risk Factors of the Equity Asset Class

- From our earlier analogy, the equity asset class balloon has the weakest tethers to the ground and also has the weakest controls that can be used to establish a price. This simply means every small wind current can set it going in different directions and we have no way to get back to course. From an equity markets perspective, this can be a number of influences, some of which are mentioned below,
    - ➤ Market Risk (Prices, Interest Rates, Foreign Exchange, Changes in Related Instruments, Volatilities etc.)

- Credit Risk (Unable to fulfil loan obligations)
- Business or Operational Risk (Information Security, Key Employees leaving, Infrastructure Issues, Security Threats etc.)
- Reputational Risk (Involvement in a lawsuit, negative press publicity etc.)
- Regulatory Risk (Changes in the legal environment that can be counterproductive)
- Industry Risk (Rapid changes in the industry can make firms obsolete)
- Liquidity (Trading volumes can go down rapidly during times of stress and the bid-ask spreads can widen)
- Earnings Risk (Variation in historical earnings, sales, dividends, uncertainty in projected dividends and future earnings)
- Size Risk (Threat of hostile takeovers or the firm might suddenly lose favour with the small number of Analysts and Investors)
- Financial Leverage (If Debt / Equity ratio is higher, the firm is more risky)
- Fundamental and Technical Risk Factors (Price / Earnings, Price/Book, Price/Sales, Industry Ratios, Historical Growth in Sales & Earnings, Expected Growth, Change in Profit Margins, Asset Turnover, Overhead Ratios, Price Momentum, Price Reversal, Earnings Momentum, etc.)
- Other Macro Factors (Sensitivity to inflation, economic growth, employment, credit spreads etc.)

- *Non-Linearity*. Apart from the above, heavily studied risk indicators, we need to be very cautious to watch out for the non-linear property of financial instruments. This is best demonstrated when attempting to get Ketchup out of a bottle onto your food. Steadily increasing hits on the base of the bottle don't yield steadily increasing amounts out the other end. None will come for a while and then a lot will. This will have great implications when we look at factor models, which are linear, in later sections. The financial markets are highly non-linear. This has been observed as huge drops in the stock market when nothing much has been happening prior to the drop.

- *The Law of Averages does not apply*. We might have observed lots of prices and other data for long periods of time and come up with probabilities. Such probabilities work in very structured environments like a casino. The financial markets due to their highly complex nature defy all odds and we need to be prepared for uncertain eventualities.

- *The assumption of normality is invalidated on many counts*. Most Pricing and Risk models assume that prices are distributed either normally or log normally. This makes modelling simpler, but we need to be mindful of the many drawbacks this holds and the situations where this assumption can breakdown.

- *No discussion about risk is complete, without a mention of Black Swans*. These are events, that are very hard to anticipate; but their effects are widely noticed and in retrospect, we seem to have known all about them. This also deals with various biases we have in our knowledge; how we interpret things and concoct explanations for things that happened in the past. Having said that, history is perhaps the best guide we have to prepare for the future. What we need to is, just be aware of the limitations in using history to predict the future.

- So far, we have talked about the unknowns that we know about. What about the unknowns that we don't know about. The only thing, we know about these *unknown unknowns* are that, there must be a lot of them, hence the need for us to be eternally vigilant.

## VI. Fundamentals of Multi-Factor Models

- We generally use Multi-Factor models for three main purposes
  - ➢ Risk Control or Management (Considered in point VII and XII below)
  - ➢ Alpha Generation (Considered in point VIII, IX and XI below)
  - ➢ Performance Attribution (Considered in point XV below)
- This is required for the portfolio construction process, for which we need to forecast the expected returns; forecast the variances in these returns and later take stock of how we performed relative to these expectations.
- Usually, multi factor models have four main components - A security's exposure to the factors, the excess returns, the attributed factor returns and the specific returns.
- The core of it is the attribution of asset returns to chosen common factor and specific returns, plus forecasts of the variances and covariances of these common factor and specific returns. Formally, we can denote this as

$$R_{it} = a_i + b_{i1}F_{1t} + b_{i2}F_{2t} + \ldots + \varepsilon_{it}$$

  *$R_{it}$ is the return on the asset i in period t*

  *$a_i$ is the intercept for asset i*

  *$F_{kt}$ is the factor k during time t*

  *$b_{ik}$ is the sensitivity of asset i's return to factor k*

  *$\varepsilon_{it}$ is the security specific (idiosyncratic) portion of the return on asset i*

- One of the more commonly used techniques is to perform multi-variate regressions either across the time series or cross-sectional across security returns to arrive at the exposure of the different stocks to various factors. This is known as arriving at the factor loadings corresponding to the factors that best explain the security returns.
- Generally we use macro-economic factors (like inflation, GDP growth, change in industrial production, spread over government bonds etc.) or fundamental factors (like firm size, dividend yield, book-to-market ratio, industry classification etc.) to understand the returns and variance structure across a universe of securities.

- To cover the non-conventional risk factors mentioned in the V section, we could build models tailored to capture that particular risk aspect. For example, to capture industry risk, we can look historically across firms in an industry, their growth rate, their rates of emergence and disappearance etc. and use the results from such a model to rank different industries or give a score for the rate of change within an industry. Such a ranking can then be used an inputs to more conventional factor models.

- We can also use other techniques like Principal component / Maximum likelihood analysis across security returns to determine the main factors.

## VII. Risk Management

The analogy of ***building a plane and flying it*** to constructing a Model and Trading with it, will help us consider the associated risks in a better way. ***Modelling would be the phase when we are building a plane, and the outcome of this process is the plane or the model which we have built; trading would then be the act of flying the plane in the turbulent skies, which are the financial markets***. The modellers would then be the scientists (also engineers) and the pilots would be traders. It is somewhat out of the scope of this document to discuss questions regarding what kind of person can be good at both modelling and trading.

### 1. From an Equity Modelling perspective

- ➢ The multi factor model will decompose overall portfolio risk and help identify the important sources of risk in the portfolio and links those sources with aspirations for active return.
- ➢ We need to use ***the right principles, the right material and the right processes***.

- The right principles would mean understanding certain concepts that determine the relevant measure of risk for any asset and the relationship between expected return and risk when markets are tending towards equilibrium. Examples for these are the *Capital Asset Pricing Model, the Arbitrage Pricing Theory* or other multi index models.
- The right material translates to having data on the security returns and choosing the relevant factors. The amount of data and factors that is available is humongous. We need to use some judgement regarding how much history to use. We also need to be attuned to Significance and Causality among the factors. All this can involve some independent data analysis.
- The right process would mean using judicious concepts from econometric / statistical theory. Some examples would be to check for the stationarity of variables, to normalize the variables to scale them properly, to see if there is any correlation between the independent variables and correcting for it (Multi-Collinearity). We need to make sure no variables that would have an impact are left out (Omitted Variable Bias)
- There needs to be *a lot of tinkering*; this means we need to have a continuous cycle of coming up with a prototype, testing how it works and making improvements based on the performance. This is especially important in the financial markets, since we are chasing moving targets, as implied by our earlier discussion on quasi-equilibriums.
- Modelling needs to be well thought out, with due regard to anticipating as many scenarios as possible and building in the relevant corrective or abortive mechanisms when adverse situations occur.
- Given that, we are never close to accomplishing a perfect model, which can

handle all cases without failure and without constant changes, we would need to constantly supervise the outcomes; hence models that are simple and robust are better suited, since it is easier to isolate the points of failure when things get rough. Robust here means producing similar results under a variety of conditions, with some changes to the inputs or the controls.

## 2. From an Equity Trading Perspective

- Trading would need a *good understanding of what the model can do and where it will fall short*.
- The model will tell us what types of risk we have in our portfolio and what returns we can expect to get from bearing that risk. Changing market conditions means the relationship between risk and return will be changing as well. But we won't know where that relationship can breakdown and what happens when some of the factors cross the boundaries within which we expect them to stay. This would mean watching out for such occurrences and recalibrating the model or making other decisions like reducing exposure to some factor etc.
- Hence, we would need to react rapidly to events as they unfold, which means we need to be able to detect events in real time with a good amount of precision. This would be like a pilot reading the many gauges on his dashboard and responding appropriately. We would need good access to market data or use as many data points as we can assemble.
- A watchful eye or sensor(s), which is able to detect our performance, a good feedback mechanism that can take corrective actions based on the inputs from the sensors. The amount of data that needs to checked and the speed with which information can change means we are better of having many automatic

> procedures that help us check the levels of various parameters.

> ➢ The feedback mechanism here can be someone manually looking at results of the model and changing either the inputs or the parameters of the model. It could also be automatic where an algorithm can detect the changes and take responsive action. ***The combination of manual effort and computer programs, man and machine working in tandem, is at present a good way to approach risk management while trading.***

> ➢ We need to check the Tracking Error of our portfolio with respect to a benchmark, measured as the volatility of the difference in returns between the portfolio and the benchmark. Passive managers want to minimize this error. Active managers want to outperform the benchmark. They need to monitor the risk results to see how they are positioned, what is their tracking error (active risk) and take on risk in areas they believe they can outperform.

> ➢ We need to be judicious about not intervening too much, since every intervention has an associated penalty or cost, not to mention, the emotional component of human involvement can cause bad results to get worse.

## VIII. Equity Investment Model and Equity Trading Strategy

- Using our earlier analogy of building a plane and flying it, we introduce an additional complexity, which will help us understand the investment modelling aspects and the corresponding trading strategy.

- Our investment decisions are made over time and so we set the direction of forward movement in time to be equivalent to flying the plane forward. Since we cannot see what will happen in the future; to fly the plane forward, we should not be able to see what is in front of us. This is equivalent to ***a plane with the front windows blackened***

- *out*. All we have are rear view mirrors (most planes don't exactly have rear view mirrors, but let us imagine our plane having one) and windows to the side.
- As we are cruising along in time, what we have with us is the historical data or the view from behind and real time data which is the view from the side, to aid in navigating our way forward or to the future.
- We use the historical data to build our model and then use the data from the present to help us make forecasts for what the future holds. ***The modelling would involve using data inputs to come up with outputs that can help us decide which securities to pick***, or to help set the direction of motion. The ***trading aspect*** would involve using the model outputs and checking if that is the direction in which we want to be heading, that ***is actually deciding which securities to pick***, and watching out for cases where the predictions are not that reliable.
- As we can see from the analogy, this is the more challenging aspect of portfolio management and the use of multi factor models.
- Capital Asset Pricing Model or Arbitrage Pricing Theory (also other models) can help us identify stocks that should be overweighed or underweighted in the portfolio. We can find securities that are cheap or expensive as given by the excess return or the alpha, relative to the model.
- ***If we are able to identify factors that explain a significant portion of security returns, or we have factors in mind that we think will bear a greater portion of the future risk and hence yield higher return, we overweight securities whose return is explained to a greater extent by those factors***.
- We can also use fundamental research techniques like the Dividend Cash Flow Model to form expectations of future security prices.
- If our performance is tied to a benchmark, we will pick securities that contribute to a

- big portion of the return in the benchmark. If our intention is to outperform the benchmark, we make intentional picks based on our forecasts of expected return that would exceed the benchmark. Based on these selections, we might be exposed to factors that are incidental or they will be the unintentional choices of our active selection. For example, if we pick securities that have high exposure to GDP growth, we may end up getting more exposure to certain sectors or industries and this may not be exactly what we want. This unintended exposure needs to be managed.

- Once we have the forecasts of expected returns, we can construct optimal portfolios that implement bets on those returns. We maximize utility, defined as risk adjusted returns depending on our level of risk aversion. The Covariance of the factors from the risk model that we discussed in the previous sections will be used here along with the expected returns. The optimization can include various constraints like liquidity, size of the firm, total exposure to a particular sector, minimum dividend yield etc. The results of this step and the revisions performed here based on market conditions would be the more trading focused aspects of investment.

- If two assets are similar, but one has a slightly higher forecasted return, traditional optimization techniques allocate everything to asset with the higher forecasted return and nothing to the other asset. This issue can be resolved by using the ***Black Litterman Model***, where we assume a prior equilibrium distribution of the assets and apply our views to tilt the weights based on the strengths we associate with those views.

- We also need to be highly sensitive to regime changes, which mean the data from the past will fail to apply to future views. This would mean, we have to go back to the drawing board and come up with new models and factors.

- The finer aspects of the distinction between trading and modelling from the earlier

sections apply here, almost verbatim.

## IX. The Model for Security Selection

We can rank the securities in the universe based on single factors (here a factor is a measure for comparing securities, e.g. Price-Earnings Ratio) or combinations of factors (we refer to this combination as a model). The different factors in a model can be equal weighted or the weights can be seeded based on an intuitive understanding of the relationships between the individual factors.

The factor weights can also be obtained as the result of an optimization process, which maximizes the returns or minimizes the variance for different groups of securities, over different historical periods. The returns can be absolute returns or the returns of one group relative to another group. These different groups of securities can be at the top or bottom quartile, or a certain percentile at the top and bottom of the ranking. We can also consider the excess returns of groups of securities versus the benchmark returns, in which case, we just go long the securities in the chosen group. We can also consider the excess returns of one group of securities versus another group, in which case we are long the first group and short the second group. Mean Variance Optimization can lead to unstable results, but this is a useful starting point.

We can also optimize the factor weights to achieve a certain ranking of securities at a certain point in time, that provides a decent return profile (either relative or absolute) for a certain top percentile of the ranking and ensure that this ranking is stable over successive investment horizons. We refer to the length of these successive investment horizons as the holding period. *The stability of the ranking can be measured by the Information Co-efficient (IC) of the ranking. This is calculated as the rank correlation (measure of the degree of* *similarity between two rankings) of the security ranking at different points in time. A higher information ratio (mean IC divided by standard deviation of IC) indicates higher stability and higher predictive ability.* The IC is generally measured as the rank correlation between the ranking of the universe based on a factor or model at a certain point in time and the ranking of the universe based on security returns at the end of the holding period. The factors weights can also be obtained through statistical methods like the Principal Component Analysis, but the results are not easily interpretable in this case.

Another useful measure is the *hit-ratio, measured as the number of holding periods over the historical back test interval over which the information co-efficient is positive*. Higher the hit ratio is, higher is the predictive ability of our model. *The hit ratio or the Information Ratio can be used to establish a confidence level for this particular view on the securities we have in this group.* For negative information ratios, we set the confidence level to zero. This will be an input to the Black Litterman model, the usage of which will be discussed in the next section.

So *our selection model boils down to picking securities or groups of securities with high relative returns and high information ratios*. Once we find factors that are showing excess returns in a group of securities, versus another group or the universe itself, we can also run regressions of these factors across the returns of the group showing excess returns to ensure there is statistical significance.

There are numerous ways in which a model can be selected and hence for the implementation, we need to have a trial and error approach and pick the more suited ones at that point in time. Extensive automation to calculate various ranks, IC's, means, regression co-efficients etc. would aid the selection process greatly. We also need to fit the model for a sample and test it across other samples. (In-Sample/Out of Sample methodology)

# X.    The Weights of Portfolio Positions

We could use any of the numerous Mean Variance Optimization (MVO) methods to arrive at the portfolio weights. There are many drawbacks with an MVO approach; the main ones being the problem of unintuitive, highly concentrated portfolios, with high sensitivity to the inputs where the estimation errors can get magnified. The Black and Litterman model is a good way to get around these issues.

To implement the Black Litterman model, which is Bayesian, we need market equilibrium expected returns as a starting point. These returns and the associated return variance would be our prior distribution. If our universe is a price weighted index like the DJIA, we need to construct a market capitalization weighted proxy for this index with all the components and use these weights to get the equilibrium expected returns as our starting point.

$$\boldsymbol{\pi = \lambda \Sigma w_{mkt}}$$ *where, π is the Implied Excess Equilibrium Return Vector (N x 1 column vector); λ is the risk aversion coefficient; Σ is the covariance matrix of excess returns (N x N matrix); and, $w_{mkt}$ is the market capitalization weight (N x 1 column vector) of the assets.*

We can then use the model for security selection outlined in the previous section and get relative views (or absolute views) on return performance for specific securities or groups of securities. We can also get the confidence level for each view we associate with a security or group of securities, by the hit ratio or the information ratio of the information co-efficient, defined in the section IX.

The variance of the error term associated with each view represents the uncertainty of the view. It can be obtained by the weight matrix that expresses our views and the co-variance matrix of our returns or a better way would be to find the weights that express full confidence in a view and tilt it so that the variance of the error term is proportional to our actual confidence in the view. We then combine our prior distribution with our views and the variance of the error term associated with each view to get a new posterior distribution. From this, the formula for the new Combined Return Vector (E[R]) is

$$E[R] = [(\tau\Sigma)^{-1} + P'\Omega^{-1} P]^{-1}[(\tau\Sigma)^{-1}\Pi + P'\Omega^{-1}Q]$$

*where, E[R] is the new (posterior) Combined Return Vector (N x 1 column vector); τ is a scalar, that is calibrated such that its value affects the variance of the error term associated with the views, but will not affect final results; Σ is the covariance matrix of excess returns (N x N matrix); P is a matrix that identifies the assets involved in the views (K x N matrix or 1 x N row vector in the special case of 1 view); Ω is a diagonal covariance matrix of error terms from the expressed views representing the uncertainty in each view (K x K matrix); Π is the Implied Equilibrium Return Vector (N x 1 column vector); and, Q is the View Vector (K x 1 column vector).*

We can then extract the posterior weights, w, that overlays our views onto the equilibrium weights as

$$w = (\lambda\Sigma)^{-1}E[R]$$

## XI. Analysis of the Portfolio Alpha and Beta

For simplicity, let us assume that the Single Index Model holds. This means that we assume that the co-movement between stocks is due to the single influence of the benchmark index (or the DJIA universe in our case). To introduce other influences that are industry specific, security specific or macro specific, would be a simple extension of this basic idea and would just involve more matrix and algebraic manipulation. With this assumption, we can decompose the return of the stock into the basic return equation as,

$$R_i = α_i + β_iR_m + ε_i$$ where, $R_i$ is the return on the security; $R_m$ is the return on the market (or universe in our case); $α_i$ is the expected value of the component of security i's return that is independent of the market's performance; $ε_i$ is the error associated with $α_i$; $β_i$ is a constant that measures the expected change in $R_i$ given a change in $R_m$.

We can estimate $α_i$ and $β_i$ by running a regression of historical stock returns against the returns of the market. Regression analysis also ensures that $R_m$ and $ε_i$ are uncorrelated over the historical time period under consideration. Beta is given by the co-variance of the security return with the market return divided by the variance of the market return. $β_i = σ_{im}/σ_m^2$. Taking expectations of the basic equation, we get,

$$E(R_i) = α_i + β_iE(R_m)$$ or equivalently, $$α_i = E(R_i) - β_iE(R_m)$$

From this, it is easy to see that to maximize alpha, we need to maximize the expected return on a security or the product of Beta and expected return on the market has to be small. This means that for a given return on the market, the Beta has to be small to maximize alpha or the security's return must have a low co-variance with the market return or a big portion of a security's return is not explained by the return on the market.

When we introduce other factors that can be used to explain the returns on a security, the same principles used above apply. To maximize alpha, we need to ensure that the expected return on the security is maximized and beta of the security return against that factor (regression co-efficient) is minimal.

## XII. Risk Evaluation in the Resultant Portfolio

Continuing with the single index model, we can derive the equations for the return variance of a security (1); return co-variance of a security with another security (2); and the variance of a portfolio (3) as below.

$$\sigma_i^2 = \beta_i^2 \sigma_m^2 + \sigma_{ei}^2 \text{ -- (1); } \sigma_{ij} = \beta_i \beta_j \sigma_m^2 \text{ -- (2); } \sigma_p^2 = \beta_p^2 \sigma_m^2 + \Sigma w_i^2 \sigma_{ei}^2 \text{ -- (3)}$$

*where, $\sigma_i^2$ is the variance of security i, $\sigma_{ij}$ is the co-variance of security i and security j, $\sigma_p^2$ is the variance of the portfolio, $\sigma_{ei}^2$ is the variance of the error term. $w_i$ is the weight of security i in the portfolio.*

To minimize the risk, we can see that the portfolio beta (which is the weighted sum of the betas of the individual securities in the portfolio) has to be minimized and the weights have to be minimized. So we can pick securities with smaller betas relative to the universe and pick at least a few of them so that the individual weights are less.

## XIII. Portfolio Rebalancing Criteria

Our main criteria for rebalancing would be to ensure that the portfolio tracking error is minimized. Tracking error is defined as the standard deviation of the active returns (difference between portfolio and the benchmark returns). As tracking error increases, we are moving away from our intended allocation weights.

For simplicity, we assume that transaction costs (taxes, market impact, commissions etc.) are linear and rebalancing benefits (related to reducing the risk of the portfolio) are quadratic in nature. As the portfolio drifts away from the intended allocations, the costs increase linearly and the benefits increase in a quadratic manner, which means, at some point, the benefits will outweigh the costs. We can use this point as the trigger point for our rebalancing. Formally, we get the ***Rebalancing Trigger Point*** as

$$\mathbf{KC_i / (\sigma_i^2 + \sigma_p^2 - 2\sigma_i\sigma_p\rho_{ip})}$$ *where, K is the risk tolerance, which needs to be calibrated separately based on the investment goals of the portfolio. $C_i$ is the transaction costs for asset i; $\rho_{ip}$ is the correlation of asset i and the portfolio*

From this, it follows that, we need to rebalance whenever the Trigger point is less than one, assuming our risk tolerance, K, captures the extent of imbalance that is acceptable between transaction costs and risk in the portfolio.

A more complex strategy would involve a ***dynamic programming approach***, which minimizes the expected cost based on an optimization involving a certain cost function for the transaction costs and a utility based cost function for the tracking error that results from holding a sub-optimal portfolio.

At time t, $w_t$ is our state or the weights in our portfolio; $u_t$ is our policy for this state, that is how we increase or decrease the weights; and $n_t$ is the state uncertainty, which is generated from the return process of the securities. The state transition can be defined by the simple multiplicative function (1), though it can be any arbitrary function.

$$\mathbf{w_{t+1} = (1 + n_t)(w_t + u_t)} \text{ -- (1)}$$

*where, $w_{t+1}$ represents the new state which is influenced by the prior state $w_t$, the action taken $u_t$, and the uncertainty in the system dynamics $n_t$.*

We write the cost functional recursively as

$$\mathbf{J_t(w_t) = E[G(w_t, u_t, n_t) + J_{t+1}(w_{t+1})]}$$

*where, G is the cost for the current period and $J_t$ is the expected future cost from t onwards given all future decisions. So, the cost at any given period is the expected cost from t to t+1 along with the expected cost from t+1 onwards.*

At each time t, the optimal strategy is to choose $u_t$ such that the cost is minimized:

$$\mathbf{J^*_t(w_t) = \min_{u_t} E[G(w_t, u_t, n_t) + J_{t+1}(w_{t+1})]} \text{ -- (2)}$$

The challenge is therefore to determine the values $J^*(w)$. This is done by simulating the price process across the desired time interval and calculating the different values of $\mathbf{J_t(w_t)}$ till we reach convergence, that is, till we reach the fixed point such that $J^*_t(w_t) = J^*_{t+1}(w_t) = J^*(w)$. The optimal rebalancing decision is to choose the policy $u^*_t$ that minimizes (2).

We specify the cost function as,

$E[G(w_t, u_t, n_t)] = \tau(u_t) + \varepsilon(w_t + u_t)$ *where, τ(u$_t$) can be a linear transaction cost depending on the adjustment, u$_t$, made to the weights; ε(·) represents the sub optimality cost due to the tracking error.* $\varepsilon(w_t + u_t) = 0$ *whenever* $w_t + u_t = w^*$*(i.e. we choose u$_t$ so that we rebalance to the target portfolio); otherwise,* $\varepsilon(\cdot) > 0$.

For any given portfolio weights, w, the expected utility from holding those positions, can be expressed as a risk-adjusted rate of return, given the risk preferences embedded in an appropriate utility function (quadratic, log, power etc.). We can then write the cost of the tracking error as the difference between the risk-adjusted rates of return associated with holding optimal or suboptimal weights. As the number of assets increases, estimation of the returns, variances and co-variances of all the assets becomes more involved. Also, the number of simulations to achieve convergence increases significantly, requiring massive use of computing power.

Other simple strategies for rebalancing include periodic rebalancing, rebalancing when the tracking error crosses a certain threshold, when the allocation weights cross a certain band around the target weights, when risk increases beyond a threshold and combinations of these strategies.

## XIV. Trade Execution and Market Impact
### 1. Equity asset class

- Here, we look at ***Transaction Cost Analysis (TCA)***, which is most developed

for equities. Drawing an analogy between "Portfolio Management" and a student studying for "An Examination", we can consider it as a three pronged process. The planning or the "Pre Trade" phase is when the student is preparing for the examination; the "Execution of Trades" constitutes the real test of one's mettle and is equivalent to the student taking the examination; and the "Post Trade" measurement of performance versus different price benchmarks becomes the Scorecard.

- Given that our current focus is Execution centric, formulation of an investment hypothesis is viewed as a separate process that precedes the actual details on how to implement a particular investment objective or an execution schedule. Going back to our Examination analogy, this is simply a student's decision regarding what to study, why study it and related aspects. Irrespective of how an investment strategy is formulated, the rest of the discussion applies to it, in its entirety. Clearly, there is a feedback loop from the Scorecard phase, where we gauge our results, to the Planning phase where we apply any lessons we learn, towards further improving our Performance.

i. **Pre Trade Metrics**

a) *Market Impact*

> Market Impact falls under the category of transaction costs incurred to exceed a certain performance benchmark by forming reasonable return expectations and controlling the risk that comes with the pursuit of the opportunities that can yield the performance targets. Broadly speaking, Market Impact is the indirect cost that occurs because of the transaction itself and is fairly independent of the commissions, taxes, exchange fees and other external costs, though it is affected by many external factors.

Ravi Kashyap                                                                                                    Page 32

- At the outset, it might seem that it would be fairly straight forward to develop a market impact cost model by observing the costs associated with previous trades. We could categorize trades into different buckets such as trade size, asset market capitalization, market etc. Then the mean of past costs for a given bucket would be a reasonable indicator of future costs. Upon closer observation, a few reasons make it clear that this method would not work.
    - Market Impact cannot be directly observed, it must be estimated. To reduce estimation error, large statistical samples are required, which implies gathering data over extended time periods. But the underlying process that creates market impact is a highly dynamic one and hence long term averages have very little forecasting ability.
    - The level of information is uneven. The most highly traded assets have very little market impact and the lightly traded assets have much higher impact. But we need to understand the higher costs and be able to predict them better since it is the assets with higher costs that are instrumental in determining the overall portfolio costs and the portfolio construction strategy.
    - Investors avoid costly trades under most possible circumstances. Hence, the data we observe is censored and does not contain many such trades. So any model that is calibrated to observed data will not perform that well under circumstances that seem to warrant higher costs.
- The Market Impact model relies on a framework relating the movement in stock prices and other auxiliary variables like Order Size, Trade Time, and Volatility etc. to Impact Costs.

- The Price Impact is decomposed into two components, the Permanent Impact and the Temporary Impact.
- The permanent component is determined by the fundamental economic forces acting in the market. This reflects the movement in price, due to the buy and sell demands on the security and hence this is independent of any timing related decisions made to buy and sell the securities.
- The temporary component occurs over the short term time horizon and is the price concession that is required to attract counterparties. Since it is a key determining factor in whether a transaction can occur successfully, it is highly sensitive to how the trades are scheduled for execution.
- Based on the parameters we can observe in the market, we define the following price points that are required to determine the impact of any order.

    $S_0$ - *Market price before order begins executing*

    $S_{Post}$ - *Market price after this order is completed*

    $S_{Avg}$ - *Average Realized price on this order*

    Permanent Impact, $\mathbf{I = (S_{Post} - S_0)/ S_0}$

    Realized Impact, $\mathbf{J = (S_{Avg} - S_0)/ S_0}$

- The Post Trade price, $S_{Post}$ should capture the permanent effects of the program. That is, it should be taken long enough after the last execution that any effects of temporary liquidity effects have dissipated. The temporary impact is defined as the realized impact minus a suitable fraction of the permanent impact.
- Asset Prices are assumed to follow an Arithmetic Brownian Motion with the drift term depending on the trade rate, v.

$$dS = S_0\, g(v)\, dt + S_0\, \sigma\, dB$$

$$S_{Avg} = S(t) + S_0\, h(v)$$

*g(v) is the permanent impact function*

*h(v) is the temporary impact function*

*S(t) is the price of the asset at time t*

*σ is the volatility of the stock*

*B(t) is a standard Brownian Motion*

*Trade Rate, v = X/T*

*X is the number of shares; it is positive for buy and negative for sell orders*

*V is the average daily volume*

*T is the total time of trading*

➢ Proceeding further with the above framework, we can assume different functional forms and include variables for bid-ask spread, shares outstanding, market capitalization, country, sector, corporate action indicators, etc. could be included in addition to the factors already considered. We can then use different data sets that contain order and execution information to calibrate the model. The unknown Greek alphabets below are then determined using advanced numerical techniques on the data set chosen for the analysis.

$$I = \sigma\, \gamma\, T\, sgn(X)\, |X/VT|^{\alpha}\, (\theta/V)^{\delta}$$

$$J - I/2 = \sigma\, \eta\, sgn(X)\, |X/VT|^{\beta}$$, sgn is the sign function.

b) **Market Risk.** The Risk inherent in any trading program determines how quickly one would like to take any trading program towards completion. The market risk is calculated as the volatility of the portfolio over the expected time or the desired time to completion. Higher the risk, the more likely is the price to drift away from the desired price used in the portfolio construction procedure and hence the desired time of completion needs to be lower. But if the time to completion decreases, we soak up a higher percentage of liquidity from the market, potentially increasing our Market Impact Costs. Hence achieving optimality in any trading program is a trade-off between the Market Risk and the Market Impact. The time to completion is the variable that will be controlled to achieve a desired end result.

c) **Tracking Error**. This will be the standard deviation of the return difference between the portfolio and the benchmark over a certain historical period.

d) **Bid Ask Spread.** This will be based on the average spread for the individual stocks for the entire trading day averaged over a certain number of trading days. The corresponding figure for the portfolio will be arrived at by using the weights of the individual securities in the portfolio.

e) **Liquidity.** This will be based on the relative size of the order and the trading volume for the entire trading day averaged over certain number of past trading days. The corresponding figure for the portfolio will be arrived at by using the weights of the individual securities in the portfolio.

f) **Beta**. We calculate beta as covariance of the asset returns versus the market, divided by the market variance. We make an adjustment to compensate for the movement of the security beta towards market beta over time using a technique developed by Blume. He corrected past betas by directly measuring the adjustment to one and assuming that the adjustment in one period is a good estimate of the adjustment in the next. This direct measurement is done by regressing Betas for a later period against the Betas for an earlier period. The corresponding Beta figure for the portfolio will be arrived at by using the weights of the individual securities in the portfolio.

  ii. **Post Trade Metrics**

  Most of the Post Trade Metrics tend to be simple comparisons of the average executed price versus different price benchmarks like the Volume Weighted Average Price, Open, Close, Previous Close, Arrival, Volume Weighted Price Over an Interval, etc.

## 2. Fixed Income Asset Class

TCA analysis in fixed income cannot be readily calculated like the equity markets, but due to the non-anonymous nature of the trading, transaction costs could be skewed towards the higher end for smaller clients.

## 3. Foreign Exchange Asset Class

TCA analysis in FX is still in the developing stages, but like the fixed income markets, due to the non-anonymous nature of the trading, transaction costs could be skewed towards the higher end for smaller clients. There are two methods that are generally used, one compares the execution price to the market price (available from centralized service providers) at the time of execution; the other compares execution price to the average of the high and low price for the day.

# XV. Portfolio Performance Measurement

- There are three basic forms of attribution. We will consider the common elements and look at how they differ for the individual asset classes.
    - *Multi-Factor analysis*. This was considered in sections VII, VIII, IX, XI and XII for the forecasting of risk and expected returns as well. It works in a similar way but seeks to attribute the observed returns on the portfolio across different factors

and the asset specific returns.

- *Style Analysis*. It will determine the investment style based on the portfolio rate of return. An example of this is the Sharpe Ratio which measures portfolio return in excess of the risk free rate divided by the standard deviation of the portfolio return. Sometimes, we need to look at the excess return achieved for a given level of risk. This would mean locating the return corresponding to the portfolio in question, on the line connecting the market portfolio (this is a hypothetical portfolio that represents all the investable assets held in proportion to their market value) to the risk free rate of return, when risk and return are plotted on a graph. We are using the standard deviation for the two metrics above. We could instead use the beta of the portfolio and we get two alternate measures called the Treynor and Jensen measure. All these categorize the risk and return metrics into different categories for easier comparisons.

- *Return Decomposition Analysis*. We attribute performance versus certain benchmarks. The active management effect is the different between total portfolio return and total benchmark return. This is the sum of three effects, Allocation, Selection and Interaction.

    o *Allocation* refers to the ability of a portfolio manager to allocate across various segments. The allocation effect determines whether the overweighting or underweighting of segments relative to a benchmark contributes positively or negatively to the overall portfolio return. Positive allocation occurs when the portfolio is overweighed in a segment that outperforms the benchmark and underweighted in a segment that underperforms the benchmark. Negative allocation occurs when the portfolio is overweighed in a segment that underperforms the benchmark

- and underweighted in a segment that outperforms the benchmark.
    - The *selection* effect measures the ability of the manager to select securities within a given segment relative to a benchmark. The over or underperformance of the portfolio is weighted by the benchmark weight, therefore, selection is not affected by the allocation to the segment. The weight of the segment in the portfolio determines the size of the effect.
    - The *interaction* effect measures the combined effect of the selection and allocation decisions within a segment. If a segment is overweighed and the selection was superior, the interaction effect is positive. If the selection was good but the segment was underweighted, the interaction effect is negative.
- We need to measure if there are any buy or sell decisions done in anticipation of movements in the markets. This is known as *Timing*. We can test for timing by running a regression of the following form. If in the below regression, the constant $c_i$ is zero then it shows that a straight line can explain the returns of the portfolio, which means there is not much timing ability.

$$\mathbf{R_{it} - R_{Ft} = a_i + b_i(R_{mt} - R_{Ft}) + c_i(R_{mt} - R_{Ft})^2 + e_{it}}$$

*$R_{it}$ is the return on the portfolio in period t*

*$R_{mt}$ is the return on a major index or the benchmark in period t*

*$R_{Ft}$ is the return on the riskless asset*

*$e_{it}$ is the residual returns of fund i, in period t*

*$a_i$, $b_i$, $c_i$ are constants.*

- We also need to look at the extent of diversification in the portfolio, as measured by how correlated the returns on the portfolio are with respect to a major index.
- There are problems with volatility as a measure of risk. Volatility is not a very

accurate measure for the investment process since it does not differentiate between upward movements and downward movements in the price. Certain other measures, which capture the difference in price at a certain point and various historical points, and the amount of time between those points, will capture the upward and downward trajectories of the price in a better way.

- To being with, we simply take the return of our portfolio over the holding periods (say one month) over the last year. We then see at how many of these, we have outperformed the benchmark.

- We obtain estimates of alpha, beta and the standard deviation of the error term by regressing fund returns on the benchmark returns. Using these, we consider a few risk adjusted measures of performance below, from among the many possibilities. We need to track changing portfolio compositions and changes in portfolio mean and variance due to this; otherwise we will get erroneous results.

- Sharpe-Ratio is the average Portfolio excess returns divided by the standard deviation of returns over the sample period. $(R_p - R_f) / \sigma_p$ *($R_p$ is portfolio return; $R_f$ is the risk free rate).*

- Treynor Measure is the portfolio excess return divided by the systematic risk (Beta) instead of total risk. $(R_p - R_f) / \beta_p$

- Jensen's Alpha is the average return of the portfolio over and above that predicted by the Capital Asset Pricing Model (CAPM). $\alpha_p = R_p - [R_f + \beta_p (R_m - R_f)]$

- Alpha Information Ratio is the alpha divided by the non-systematic risk of the portfolio (standard deviation of the error term). $\alpha_p / \sigma_{\varepsilon p}$

- We calculate the t-statistic for the alpha estimate $(\alpha_p * \sqrt{N})/ \sigma_{\varepsilon p}$ (*N is the number of time periods; $\sigma_{\varepsilon p}$ is standard deviation of the error term*), to see what level of significance we can ascribe to the alpha. The more frequent our sampling frequency, the more accurate will be the t-statistic.

- We can see the effect of certain investment styles on performance by regressing portfolio returns on the returns across style portfolios formed based on certain investment styles like Small Cap, Large Cap, High P/E (growth) etc.

- The active management effect is the different between total portfolio return and total benchmark return. This is the sum of three effects, Allocation, Selection and Interaction. In the timing point above, cash and equities would be our two asset classes. Though this does not apply to our specific case, it is useful to do this attribution. Allocation refers to the ability of a portfolio manager to allocate across various segments. The selection effect measures the ability of the manager to select securities within a given segment relative to a benchmark. The interaction effect measures the combined effect of the selection and allocation decisions within a segment.

1. **Equity asset class**

   For Equities, we could attribute the returns to fundamental factors, macro factors, sectors, etc., that is to the various benchmarks. This would give us the excess return or the alpha. These factors and the attribution models are discussed in detail in sections VII, VIII, IX, XI, XII and in the common performance section above.

2. **Fixed Income asset class**

   Fixed income attribution is not as standardized as the equity evaluation process. We need to also look at different effects when measuring performance. We look at a set of methods that work well for fixed income instruments and follow a three step process.

- The first step is the calculation of the return for each security in the portfolio and the benchmark for each day and decomposition of them into various risk components.
- Second, the single components are aggregated according to the investment process.
- Third, the attribution effects are calculated. We look at three main effects,
  - *Carry Effect* is due to the impact of time on the returns of the instrument. This can be further decomposed into systematic and specific carry return. Systematic return comes from the reference yield curve and specific carry return is related to the spread of this security. We also have the coupon effect and the convergence effect, which as the name implies are based on the coupons payments and convergence explains the price changes that arise from the pull to par of a bond.
  - *Yield Curve Effect* is due to the yield curve changes. We can categorize the changes into parallel shifts of the yield curve; rotation of the yield curve; change in the shape of the yield curve and shortening of the bond's remaining time to maturity.
  - *Spread Effect* is the difference in yield of a security versus the risk free rate for that security (government bonds). The spread changes arise due to the credit rating on the firm, the spread due to the sector of the instrument, spread due to the specific country.
- In addition, we need to look at various auxiliary measures such as,
  - Price effect is the difference between the portfolio valuation (done in - house) and the benchmark (valuation by the index provided).

- o Trading effect shows the difference between the price at trade time and the in house end of day valuation.
- o Allocation effect, as we discussed earlier, shows the result from weighting a certain sector or category with respect to the benchmark.
- o Duration effect is due to the modified duration of the security.
- o Convexity effect, which is due to the convexity of the yield curve for the security or that particular segment.
- ➢ Principal component Analysis is a popular tool used for many yield curve related attributions.

# 3. Commodities and FX

- ➢ The returns on commodities need to be treated slightly differently due to the physical nature of the instruments and since they are primarily held for consumption. There is a cost associated with respect to storing the instrument and we need to factor this in our return attribution mechanism. We consider the case of commodity futures in detail below.
- ➢ Due to the storage aspects, commodities have something known as a *convenience yield* which reflects the market's expectations concerning the future availability of the commodity. The greater the possibility that shortages will occur, the higher the convenience yield. If users of the commodity have high inventories, there is little chance of shortage and the convenience yield will be low and vice versa. This arises because users of a consumption commodity may feel that the ownership of the physical commodity provides benefits that are not obtained by holders of the futures contracts.
- ➢ An investor rarely holds a commodity futures contract through to maturity since his intention is not to take delivery of the physical commodity but just to

- get exposure to changes in its price. In order to avoid taking delivery, the investor will close out or sell the futures contract and initiate a new long position in another futures contract that has a later maturity date.
- The returns from passively buying futures contracts and continuously rolling them forward are known as futures only returns. This return is also known as excess returns since this is earned in addition to any returns from the collateral.
- The total return is a combination of the *futures only return* and the *collateral return*.
- We split the futures only return into the *spot yield* and *roll yield*. The spot yield arises due to changes in prices of nearby contracts (in terms of maturity). The roll yield arises due to rolling a position in a sloping forward curve environment.
- The roll yield equals the convenience yield minus the cost of storing the commodity and also financing its purchase.
- In addition, we can borrow from the earlier discussion on equity and fixed income to understand the price drivers for commodities.
- *Currency Performance Attribution*. When we are looking at portfolios with multiple currency or foreign currency denominated instruments, we need to isolate the returns into the price appreciation of the assets in the local currency; the returns from converting the local asset values back to the base currency; and the cross product returns, which arise from repatriating local profits back to the base currency or the combination of the local returns and the currency return.

$$R_n = L_n + E_k + L_n E_k$$

*$R_n$ is the return in the base currency of asset n*

*$L_n$ is the local return of asset n, denominated in currency k*

*$E_k$ is the performance of currency k relative to the base currency*

It is useful to work in terms of excess returns defined relative to the risk-free rates. If we let $R_{Fk}$ be the risk free return of currency k and, $R_{base}$ be the corresponding risk-free for the base currency, the above equation can be written as,

$$R_n - R_{base} = (L_n - R_{Fk}) + (R_{Fk} + E_k + R_{Fk} E_k - R_{base}) + (L_n - R_{Fk}) E_k$$

This result decomposes the base excess return into *local excess return*, which answers the question of whether the local asset outperforms the local risk-free rate.

The second component is the *currency excess return*, which answers whether holding cash in currency k outperforms holding cash in the base currency. A positive exposure to a currency that outperforms the base currency contributes positively to the portfolio currency effect. The net currency effect is due to both holding cash and risky assets.

The third component, *cross product,* arises from the conversion of local excess profits into the base currency. This is usually a small component and is significant only if there are large local excess returns and large exchange-rate fluctuations. This does not arise from any active management decisions, but is simply a combined effect of the local investment and the currency exposure decisions.

➢ Another key concept here is that of *currency overlay techniques*, which are used to manage the currency exposure and sometimes also benefit from it. The exposures of all the foreign currencies are combined and managed separately from the assets. The exposure can be passively hedged using forward currency contracts, currency swaps, futures or options. The exposure can also be actively managed with timing decisions based on views of exchange rate performance. Here a decision needs to be made and constantly revised regarding the ratio of the total currency exposure that needs to be active managed versus passively hedged. Here again, views can be formed based on different types of models that use fundamental factors (interest rates, balance of payments, capital flows, etc.); technical factors (based on price history); trading factors (based on options or interest rate spreads etc.) or combinations of the three. Depending on the choice of the exposure management strategy, further attribution of the currency return into these factors might be necessary.

# XVI. Conclusion

With the foundation provided by the Uncertainty Principle of the Social Sciences, we looked at the fundamental characteristics of the Equity Asset class and how trading strategies can be formulated. Once these strategies are formed, we delved into the Portfolio Construction, Portfolio Implementation, Risk Management, Rebalancing and Performance Measurement of the resulting Portfolio. We also established a way to systematically form a confidence level for any views we formulate in the Black Litterman model, as part of deriving the portfolio weights. Finally, we looked at the trading costs associated with implementing an equity portfolio with a brief foray into the implementation costs for other asset classes.

# XVII. References

Alexander, C. (1999). Optimal hedging using cointegration. Philosophical Transactions of the Royal Society of London. *Series A: Mathematical, Physical and Engineering Sciences, 357*(1758), 2039–2058. http://dx.doi.org/10.1098/rsta.1999.0416

Almgren, R., & Neil, C. (2001). Optimal execution of portfolio transactions. *Journal of Risk, 3*, 5–40.

Avellaneda, M., & Sasha, S. (2008). High-frequency trading in a limit order book. *Quantitative Finance, 8*(3), 217–224. http://dx.doi.org/10.1080/14697680701381228

Bodie, K., & Kane, A. (2002). Marcus. *Investments*.

Brealey, R. A., Stewart, C. M., & Franklin, A. (2006). *Corporate finance* (Vol. 8). McGraw-Hill/Irwin.

Caouette, J. B., Edward, I. A., & Paul, N. (1998). *Managing credit risk: the next great financial challenge* (Vol. 2). John Wiley & Sons.

Chiu, J., Lukman, D., Modarresi, K., & Velayutham, A. (2011). High-frequency trading. *Standford, California, US: Stanford University*.

Chlistalla, M., Speyer, B., Kaiser, S., & Mayer, T. (2011). High-frequency trading. *Deutsche Bank Research*.

Christodoulakis, G. A. (2002). *Bayesian Optimal Portfolio Selection: the Black-Litterman Approach*. Unpublished paper. Retrieved from http://www. staff. city. ac. uk/~gchrist/Teaching/QAP/optimalportfoliobl. Pdf

Copeland, L. S. (2008). *Exchange rates and international finance*. Pearson Education.

Dempster, M. A. H., & Jones, C. M. (2001). A real-time adaptive trading system using genetic programming. *Quantitative Finance, 1*(4), 397–413. http://dx.doi.org/10.1088/1469-7688/1/4/301

Derman, E., & Paul, W. (2009). *The Financial Modelers' Manifesto*. Retrieved from http://ssrn. com/abstract

Elton, E. J. (2009). *Modern portfolio theory and investment analysis*. John Wiley & Sons.

Engle, R. F., & Clive, W. J. G. (1987). Co-integration and error correction: representation, estimation, & testing. *Econometrica: Journal of the Econometric Society,* 251–276. http://dx.doi.org/10.2307/1913236

Fama, E. F., & James, D. M. (1973). Risk, return. & equilibrium: Empirical tests. *The Journal of Political Economy*, 607–636. http://dx.doi.org/10.1086/260061

Ferguson, N. (2008). *The ascent of money: A financial history of the world*. Penguin.

Gladwell, M. (2006). *The tipping point: How little things can make a big difference*. Hachette Digital, Inc.

Gladwell, M. (2009). *Outliers: The story of success*. Penguin UK.

Guilbaud, F., & Huyen, P. (2013). Optimal high-frequency trading with limit and market orders. *Quantitative Finance, 13*(1), 79–94. http://dx.doi.org/10.1080/14697688.2012.708779

Gujarati, D. N. (1995). Basic econometrics (3rd ed.).

Hayek, F. A. (2009). *The road to serfdom: Text and documents–The definitive edition*. University of Chicago Press.

He, G., & Robert, L. (2002). The intuition behind Black-Litterman model portfolios.

Hull, J. (2010). *Options, Futures., & Other Derivatives, 7/e (With CD)*. Pearson Education India.

Levitt, S. D., & Stephen, J. D. (2006). *Freakonomics Rev Ed LP: A Rogue Economist Explores the Hidden Side of Everything*. HarperCollins.

Mallaby, S. (2011). *More money than god: Hedge funds and the making of the new elite*. Bloomsbury Publishing.

Marx, K., & Friedrich, E. (2012). *The communist manifesto*. Yale University Press.

McCraw, T. K. (1997). *Creating modern capitalism: How entrepreneurs, companies and
Ravi Kashyap                                                                                                                       Page 47